\documentclass[letterpaper,12pt]{article}

\usepackage{graphicx}
\usepackage{float}
\usepackage{tikz}
\usepackage{amsmath,amssymb,amsthm,mathtools}
\usepackage{booktabs}
\usepackage{array}
\usepackage{enumitem}
\usepackage{xcolor}
\usepackage{url}
\usepackage{comment}
\usepackage[letterpaper,margin=1in]{geometry}
\usepackage{setspace}
\usepackage[authoryear,round]{natbib}
\bibpunct{(}{)}{;}{a}{,}{,}
\usepackage[hidelinks]{hyperref}
\usepackage{csquotes}
\input{references.tex}

\newtheorem{theorem}{Theorem}
\newtheorem{proposition}{Proposition}
\newtheorem{lemma}{Lemma}
\numberwithin{equation}{section}
\newcommand{\one}{\mathbf 1}
\newcommand{\Exp}{\mathbb E}
\newcommand{\welfarepoint}[1]{\mathsf{P}_{#1}}
\newcommand{\Law}{\mathcal L}

\title{Privacy-Preserving Disclosure and Welfare}
\author{Keita Kuwahara\thanks{Graduate School of Economics, The University of Tokyo, 7-3-1 Hongo, Bunkyo-ku, Tokyo 113-0033, Japan. Email: \href{mailto:kuke0303@g.ecc.u-tokyo.ac.jp}{\nolinkurl{kuke0303@g.ecc.u-tokyo.ac.jp}}.}
\quad Sho Mikami\thanks{Graduate School of Economics, The University of Tokyo, 7-3-1 Hongo, Bunkyo-ku, Tokyo 113-0033, Japan. Email: \href{mailto:mikami-sho910@g.ecc.u-tokyo.ac.jp}{\nolinkurl{mikami-sho910@g.ecc.u-tokyo.ac.jp}}.}}
\date{November 22, 2026}

\begin{document}
\maketitle

\begin{abstract}
We consider a monopolist selling to multiple buyers. A platform observes buyers' tastes and can publicly disclose information about those tastes. However, the platform faces a privacy constraint: the disclosure must reveal nothing about any individual buyer's taste. We show that the same pairs of consumer and producer surplus are attainable with or without this constraint. For each such pair, we construct a privacy-preserving disclosure rule under which a posted-price mechanism attains the pair and no other feasible selling mechanism earns more.
\end{abstract}


\clearpage
\section{Introduction}

Information about buyers' willingness to pay enables a monopolist to price discriminate and affects the division of surplus between consumers and the seller. However, disclosing such information may conflict with the protection of individual privacy. This paper asks whether requiring public signals to reveal nothing about any individual buyer's willingness to pay restricts the attainable welfare outcomes. The key observation is that a public signal can reveal relationships among buyers' valuations without revealing anything about any individual buyer's valuation.

We consider a monopolist selling a homogeneous good to multiple buyers, with no supply constraints and zero marginal cost. Each buyer privately knows her valuation, and valuations are independently distributed across buyers. A platform commits to and announces a disclosure rule before observing the buyers' valuations, then publishes a signal according to that rule. The signal must, however, be statistically independent of each buyer's valuation. After observing the signal, the seller commits to a mechanism that maximizes conditional expected revenue subject to incentive compatibility and individual rationality.

Our main result is that this privacy constraint does not reduce the set of attainable consumer and producer surplus pairs. The seller can combine the public signal with truthful reports from other buyers to infer information about a buyer's valuation. This allows the seller to offer each buyer a price that does not depend on that buyer's own report. For each attainable surplus pair, we construct a privacy-preserving disclosure rule under which a posted-price mechanism attains the pair. We show that the mechanism maximizes the seller's revenue among all feasible selling mechanisms under that rule.

The attainable welfare region is the Minkowski sum of the individual buyers' surplus triangles. For bounded valuations, each buyer's triangle has the form characterized by Bergemann, Brooks, and Morris (2015). Specifically, consumer surplus must be nonnegative, producer surplus must be at least the sum of the individual monopoly revenues without disclosure, and total surplus cannot exceed the gains from efficient trade. For bounded valuations without privacy constraints, every point in this region can be attained by applying the single-buyer constructions separately to each buyer.\footnote{Independent segmentations preserve the independence of buyers' valuations conditional on the disclosed segments. Averaging each buyer's allocation and transfer over the other buyers' reports yields a single-buyer mechanism that preserves expected revenue, incentive compatibility, and individual rationality. Thus, the monopoly revenue for each buyer in the disclosed segment bounds the revenue obtainable from that buyer. With no supply constraint, the seller can apply the selected optimal pricing rule for each buyer simultaneously. Under the corresponding independent segmentations, these pricing rules attain the target surplus pair.} Our results also cover unbounded values.

\subsection{Related literature}

\citet{BBM2015} characterize the surplus triangle attainable through segment-specific pricing for a given distribution of valuations. \citet{ShenTangZeng2018} provide explicit constructions of market segmentations that attain the two vertices of this triangle with minimum producer surplus. \citet{MunagalaShenXu2026} show that restrictions on the prices that can be charged in each segment can shrink the attainable welfare region. These studies allow market segmentation to change the conditional distribution of valuations and consider a seller who sets an optimal price in each segment. Relatedly, \citet{IchihashiSmolin2025} study information provision to a seller who has private information about a buyer's binary valuation and show that the attainable welfare region can be smaller than the region attainable when the designer observes the seller's information.

We consider a different information environment by imposing a privacy constraint requiring that the public signal be independent of each buyer's valuation. Moreover, rather than restricting the seller to posted pricing, we allow general incentive-compatible and individually rational mechanisms. Despite these differences, we show that the entire surplus triangle remains attainable. Furthermore, our constructed posted-price mechanisms maximize the seller's revenue among all feasible selling mechanisms under the same disclosure rule.

Our paper also contributes to the literature on information design under privacy constraints. \citet{StrackYang2024} characterize signals that are independent of protected information. In their application to price discrimination, buyers' group membership is kept private, while information about their willingness to pay may be disclosed. They study revenue-maximizing market segmentation under this privacy constraint. \citet{StrackYangBuyer2024} consider a model in which the seller and the buyer each possess a private signal about the buyer's valuation. They show that requiring the buyer's signal to be independent of the seller's signal does not reduce the attainable welfare region. \citet{FallahJordanMakhdoumiMalekian2024} study price discrimination under a privacy constraint that probabilistically masks consumers' market segments. They show that the attainable welfare region can differ from the surplus triangle attainable without privacy constraints. In contrast to these studies, we require the public signal to be independent of each buyer's valuation itself, while allowing the price offered to each buyer to depend on this signal and other buyers' reports.

The information structure used in our paper is also related to the notion of perfect sample privacy, which requires the disclosed signal to be independent of each individual sample but not necessarily of the entire dataset \citep{RassouliRosasGunduz2020}. In the proof of their Theorem~5, they construct a privacy-preserving signal by taking the sum of discretized quantiles modulo the number of intervals. Similar modular constructions appear in \citet[Appendix~A.2]{RenouTomala2012} and \citet[Proposition~1]{HeSandomirskiyTamuz2026}. Furthermore, \citet[Lemma~1]{Ichihashi2021} studies the division of an information structure into separately purchasable pieces of data that individually reveal no information about a common state but collectively reveal information about that state. We use such information-masking and recovery techniques for optimal pricing under a public signal that conceals individual valuations.

In our model, buyers' valuations are independent ex ante but may become correlated conditional on the public signal. In this respect, our paper is related to \citet{CremerMcLean1988}, who establish full surplus extraction by exploiting correlation among buyers' valuations, and \citet{McAfeeReny1992}, who show that information rents can be made arbitrarily small in more general environments. The key difference is that correlation in our model is induced by the platform's disclosure policy rather than being a feature of the prior distribution.

Finally, our paper is related to research on implementation that combines information disclosure with buyers' reports. \citet{Kraehmer2020} shows that secret randomization of information structures and consistency checks on reports can achieve full surplus extraction under certain conditions. \citet{Zhu2023} shows that individually uninformative private signals can be combined to recover the state under certain conditions, allowing the designer to achieve the same payoff as if the state were directly observable. In contrast, we achieve the desired surplus pairs through posted prices contingent on the public signal and other buyers' reports, without relying on off-path punishments.

\section{Model}\label{sec:model}

\subsection{Environment}

A monopolist sells a homogeneous good to $n\ge2$ buyers. Each buyer demands at most one unit. The seller faces no supply constraints, and the marginal cost of production is zero. The seller and buyers are risk neutral and have quasilinear utility.

Buyer $i$ privately observes her value $v_i\ge0$. The values $v=(v_1,\ldots,v_n)$ are mutually independent. Each $v_i$ has a continuous cumulative distribution function (CDF) $F_i\colon[0,\infty)\to[0,1]$ with $F_i(0)=0$, and $0<\Exp[v_i]<\infty$.

For buyer $i$, the continuous revenue function $p[1-F_i(p)]=p\Pr(v_i>p)$ is zero at $p=0$ and positive at some $p>0$ because $\Exp[v_i]>0$. Revenue tends to zero as $p\to\infty$, since revenue is bounded above by $\Exp[v_i\one\{v_i>p\}]\to0$. The maximum $r_i\coloneqq\max_{p\ge0}p[1-F_i(p)]$ is the seller's monopoly revenue from that buyer. We assume that the maximizing price $p_i^*$, called the monopoly price for buyer $i$, is unique.\footnote{Many commonly used distributions satisfy this assumption, including uniform distributions on nonnegative intervals, exponential distributions, and beta distributions.}

\subsection{Information and timing}

Before observing $v$, the platform, acting as the information designer, commits to and announces an information structure $(S,\sigma)$, where $S$ is a standard Borel signal space and $\sigma\colon V\to\Delta(S)$ is a measurable kernel, with $V=[0,\infty)^n$ and $\Delta(S)$ the set of Borel probability measures on $S$. After observing $v$, the platform draws and publishes $s\sim\sigma(\cdot\mid v)$. Write $\perp$ for statistical independence under the joint law induced by the prior and $\sigma$, $\Law(y)$ for the law of $y$, and $\Law(y\mid z)$ for a regular conditional law of $y$ given $z$. The structure is \emph{privacy-preserving} if
\begin{equation}\label{eq:privacy}
s\perp v_i\;\; (i=1,\ldots,n).
\end{equation}
The seller observes $s$ and commits to a mechanism that maximizes expected revenue conditional on the signal. Each buyer then observes the mechanism and $(s,v_i)$ and chooses how to act. The mechanism determines allocations and transfers. Each buyer's outside option gives zero utility.

\subsection{Mechanisms and incentives}\label{sec:mechanisms}

A direct mechanism specifies each buyer's allocation probability $x_i(s,\hat v)\in[0,1]$ and expected transfer $t_i(s,\hat v)\in\mathbb R$. Both are jointly measurable in the signal $s$ and report vector $\hat v\in V$. Buyer $i$'s utility is $v_i x_i(s,\hat v)-t_i(s,\hat v)$. The subscript $-i$ omits buyer $i$; thus $(\hat v_i,v_{-i})$ replaces only component $i$ of $v$ by $\hat v_i$.

By the revelation principle, we may restrict attention to incentive-compatible and individually rational direct mechanisms. Throughout, we assume that buyers play the truthful-reporting equilibrium of the direct mechanism.

For each $i$, incentive compatibility (IC) and individual rationality (IR) require the following inequalities for every $(s,v_i)$ in a single measurable set $A_i\subset S\times[0,\infty)$ of probability one and every report $\hat v_i\ge0$. The conditional expectations in these inequalities must be well defined under a fixed regular conditional law of $v_{-i}$ given $(s,v_i)$.
\begin{align}
&\Exp[v_i x_i(s,(v_i,v_{-i}))-t_i(s,(v_i,v_{-i}))\mid s,v_i]\notag\\
&\hspace{1cm}\ge\Exp[v_i x_i(s,(\hat v_i,v_{-i}))-t_i(s,(\hat v_i,v_{-i}))\mid s,v_i],\label{eq:incentive}\\
&\Exp[v_i x_i(s,(v_i,v_{-i}))-t_i(s,(v_i,v_{-i}))\mid s,v_i]\ge0.\label{eq:participation}
\end{align}
A mechanism is \emph{feasible} if it satisfies IC \eqref{eq:incentive}, IR \eqref{eq:participation}, and absolute integrability of transfers under truthful reporting\footnote{Absolute integrability ensures finite expected transfers and, with finite mean values, finite expected surpluses. The incentive and participation constraints alone do not imply absolute integrability.}:
\[
\Exp\left[\sum_{i=1}^n\left|t_i(s,v)\right|\right]<\infty.
\]
A feasible mechanism $(x,t)$ is \emph{seller-optimal} if, for every feasible mechanism $(\tilde x,\tilde t)$,
\[
\Exp\left[\sum_{i=1}^n t_i(s,v)\,\middle|\,s\right]
\ge
\Exp\left[\sum_{i=1}^n \tilde t_i(s,v)\,\middle|\,s\right]
\]
for almost every $s$.

In a posted-price mechanism, each buyer $i$ faces a measurable nonnegative price that may depend on $(s,\hat v_{-i})$ but not on her own report. She receives one unit and pays that price if her report is at least the price; otherwise, she receives no unit and pays nothing. By reporting truthfully, the buyer purchases exactly when the price is no greater than her value and thus maximizes her payoff.

\subsection{Welfare}

For a given $(S,\sigma)$ and a selected seller-optimal mechanism $(x,t)$, consumer surplus is $u$ and producer surplus is $\pi$:
\begin{equation}\label{eq:welfare}
u\coloneqq\Exp\left[\sum_{i=1}^n(v_i x_i(s,v)-t_i(s,v))\right],\;
\pi\coloneqq\Exp\left[\sum_{i=1}^n t_i(s,v)\right].
\end{equation}
The no-disclosure monopoly revenue and efficient total surplus are
\begin{equation}\label{eq:benchmark}
\pi^*\coloneqq\sum_{i=1}^n r_i,\;
w^*\coloneqq\sum_{i=1}^n\Exp[v_i].
\end{equation}
At $p_i^*$, the sale probability is positive and atomlessness gives $\Exp[v_i]\ge\Exp[v_i\one\{v_i>p_i^*\}]>p_i^*\Pr(v_i>p_i^*)=p_i^*[1-F_i(p_i^*)]$. Summing over buyers yields $0<\pi^*<w^*$.

Without disclosure, independence lets us average each buyer's allocation and transfer over the other buyers' reports, preserving expected revenue, IC, and IR. In the resulting single-buyer problems, the standard envelope argument \citep{Myerson1981} shows that posting the monopoly price maximizes revenue.

\begin{samepage}
\section{The Surplus Triangle}\label{sec:result}

\begin{theorem}\label{thm:exact}
There exist a privacy-preserving information structure $(S,\sigma)$ and a seller-optimal mechanism $(x,t)$ yielding consumer surplus $u$ and producer surplus $\pi$ if and only if
\begin{equation}\label{eq:exacttriangle}
u\ge0,\qquad \pi\ge\pi^*,\qquad u+\pi\le w^*.
\end{equation}
Moreover, every such surplus pair can be attained with a seller-optimal posted-price mechanism.
\end{theorem}
\end{samepage}

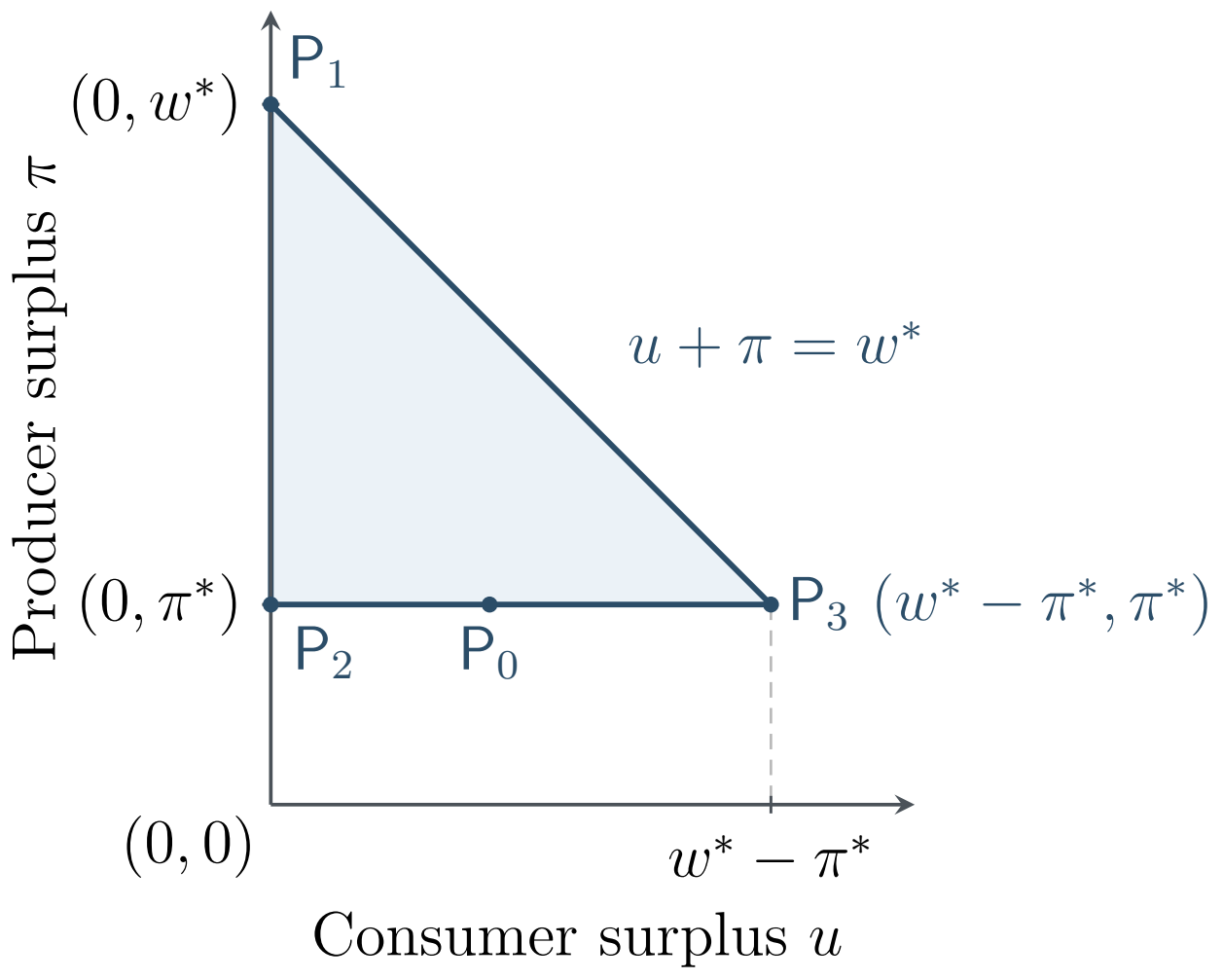
\begin{figure}[H]
\centering
\begin{tikzpicture}[x=.85cm,y=.85cm,>=stealth,font=\normalsize]
  \definecolor{welfareink}{RGB}{43,77,104}
  \definecolor{welfareshade}{RGB}{235,242,247}
  \definecolor{welfareaxis}{RGB}{75,82,89}

  \coordinate (L) at (0,1.6);
  \coordinate (U) at (0,5.6);
  \coordinate (E) at (4,1.6);
  \coordinate (N) at (1.75,1.6);

  \fill[welfareshade] (L)--(U)--(E)--cycle;
  \draw[welfareink,line width=1.1pt,line join=round] (L)--(U)--(E)--cycle;

  \draw[->,welfareaxis,line width=.7pt] (0,0)--(5.15,0);
  \draw[->,welfareaxis,line width=.7pt] (0,0)--(0,6.35);
  \draw[gray!55,densely dashed,line width=.5pt] (E)--(4,0);
  \draw[welfareaxis,line width=.55pt] (-.07,1.6)--(.07,1.6);
  \draw[welfareaxis,line width=.55pt] (-.07,5.6)--(.07,5.6);
  \draw[welfareaxis,line width=.55pt] (4,-.07)--(4,.07);

  \node[anchor=north east,inner sep=1.2pt] at (-.06,-.03) {$(0,0)$};
  \node[anchor=east,inner sep=2pt] at (-.14,1.6) {$(0,\pi^*)$};
  \node[anchor=east,inner sep=2pt] at (-.14,5.6) {$(0,w^*)$};
  \node[anchor=north,inner sep=2pt] at (4,-.15) {$w^*-\pi^*$};
  \node[rotate=90,anchor=center] at (-1.85,3.175) {Producer surplus $\pi$};
  \node[anchor=north] at (2.45,-.68) {Consumer surplus $u$};

  \node[text=welfareink,anchor=west] at (2.68,3.68) {$u+\pi=w^*$};
  \node[text=welfareink,anchor=west,inner sep=3pt] at (E) {$\welfarepoint{3}\ (w^*-\pi^*,\pi^*)$};
  \node[text=welfareink,anchor=south west,inner sep=3pt] at (U) {$\welfarepoint{1}$};
  \node[text=welfareink,anchor=north west,inner sep=4pt] at (L) {$\welfarepoint{2}$};
  \node[text=welfareink,anchor=north,inner sep=4pt] at (N) {$\welfarepoint{0}$};
  \foreach \p in {L,U,E,N} {\fill[welfareink] (\p) circle (1.55pt);}
\end{tikzpicture}
\caption{The aggregate attainable welfare region. The position of $\welfarepoint{0}$ is schematic.}
\label{fig:triangle}
\end{figure}

Figure~\ref{fig:triangle} shows the surplus triangle in Theorem~\ref{thm:exact}: the Minkowski sum of buyer-specific surplus triangles.

Without disclosure, posting the prior monopoly prices yields point $\welfarepoint{0}$, with consumer surplus $\sum_i\Exp[\max\{v_i-p_i^*,0\}]$ and producer surplus $\pi^*$. Point $\welfarepoint{0}$ can coincide with $\welfarepoint{3}$ if trade is already efficient.
At $\welfarepoint{1}=(0,w^*)$, all available surplus accrues to the seller: the mechanism achieves full surplus extraction. At $\welfarepoint{2}=(0,\pi^*)$, the seller earns the no-disclosure benchmark revenue $\pi^*$, and buyers receive no surplus. Trade is inefficient at this vertex because $\pi^*<w^*$. At $\welfarepoint{3}=(w^*-\pi^*,\pi^*)$, trade is efficient and buyers receive the residual surplus after the seller earns the same revenue $\pi^*$.

\subsection{Disclosure and optimal posted prices}\label{sec:conversion}

\begin{samepage}
To construct the disclosure rule, choose for each buyer $i$ a measurable map $g_i\colon[0,\infty)\to[0,1)$ such that the \emph{key} $k_i=g_i(v_i)$ is uniform:
\begin{equation}\label{eq:keys}
k_i\coloneqq g_i(v_i),\qquad k_i\sim \operatorname{Unif}[0,1).
\end{equation}
The keys $k_1,\ldots,k_n$ are independent because $v_1,\ldots,v_n$ are independent. For a real number $z$, define $z\bmod1\coloneqq z-\lfloor z\rfloor\in[0,1)$, where $\lfloor z\rfloor$ is the greatest integer less than or equal to $z$. With two buyers, the public signal is $s=(k_1+k_2)\bmod1$. If buyer 2 reports $\hat v_2$, the seller can compute $g_2(\hat v_2)$ and form $(s-g_2(\hat v_2))\bmod1$. When buyer 2 reports truthfully, the computed value equals buyer 1's key. The seller can offer buyer 1 a monopoly price for the conditional distribution of buyer 1's value given the recovered key, without using buyer 1's own report.

In general, the platform publishes the sum of all buyers' keys modulo one:
\begin{equation}\label{eq:mod}
s\coloneqq\left(\sum_{i=1}^n k_i\right)\bmod1.
\end{equation}
\end{samepage}
Fix $i$; since $n\ge2$, we may choose $j\ne i$. Conditional on $v_{-j}$, the signal in \eqref{eq:mod} is a fixed shift of $k_j$ modulo one. Since $k_j$ is uniform and independent of $v_{-j}$, $\Law(s\mid v_{-j})=\operatorname{Unif}[0,1)$ almost surely. As $v_i$ is a component of $v_{-j}$, $\Law(s\mid v_i)=\operatorname{Unif}[0,1)$ almost surely, so $s\perp v_i$. Fix a regular conditional distribution $Q_i(k)\coloneqq\Law(v_i\mid k_i=k)$.\footnote{The seller need not know how the disclosure rule is represented in terms of keys. The prior and the announced rule determine $\Law(v_i\mid s,v_{-i})$ up to null sets. This distribution equals $Q_i(k_i)$ almost surely: given $v_{-i}$, the signal and the key determine each other, and $v_{-i}$ is independent of $(v_i,k_i)$. The seller can thus evaluate the conditional distribution at the other buyers' reports and choose a monopoly price for that distribution.} The map $g_i$ can assign the same key to different values.

\begin{samepage}
For any probability distribution $Q$ on $[0,\infty)$, its monopoly revenue is
\[
\rho(Q)\coloneqq\sup_{p\ge0}pQ([p,\infty)).
\]
A price attaining this supremum is a monopoly price for $Q$.
\end{samepage}

Call a distribution countably supported if it assigns probability one to a countable set. The next proposition bounds expected revenue conditional on the public signal. The bound is the sum, across buyers, of their expected monopoly revenues conditional on their respective keys.

\begin{proposition}\label{prop:keyconversion}
\begin{samepage}
For each buyer $i$, suppose her conditional value distribution $Q_i(k)$ is countably supported for almost every $k\in[0,1)$, and let $p_i\colon[0,1)\to[0,\infty)$ be a measurable function selecting a monopoly price for $Q_i(k)$ at almost every $k$. Under the public signal $s$ in \eqref{eq:mod}, consider the mechanism offering each buyer $i$ the price $p_i(\kappa_i)$, where
\begin{equation}\label{eq:decode}
\kappa_i\coloneqq\left(s-\sum_{j\ne i}g_j(\hat v_j)\right)\bmod1.
\end{equation}
This posted-price mechanism is seller-optimal among all feasible mechanisms.
\end{samepage}
\end{proposition}

\begin{proof}
\textit{Reduction to a single-buyer problem.}
Fix a competing feasible mechanism $(x,t)$ and buyer $i$, and let $s_{-i}=(\sum_{j\ne i}g_j(v_j))\bmod1$. Since the values are independent and $k_i$ depends only on $v_i$ and $s_{-i}$ only on $v_{-i}$, the pairs $(v_i,k_i)$ and $(v_{-i},s_{-i})$ are independent. Checking probabilities on rectangles gives the conditional law of $(v_i,v_{-i})$ given $(k_i,s_{-i})$ as the product of the law of $v_i$ given $k_i$ and the law of $v_{-i}$ given $s_{-i}$. Given $k_i$, the signals $s$ and $s_{-i}$ determine each other. Substituting $s_{-i}=(s-k)\bmod1$ when $k_i=k$ gives, for almost every $(s,k)$,
\[
\Law(v_i,v_{-i}\mid s,k_i=k)
=Q_i(k)\otimes\Law(v_{-i}\mid s_{-i}=(s-k)\bmod1).
\]
Conditioning on $(s,v_i)$ is likewise equivalent to conditioning on $(s_{-i},v_i)$ because $k_i=g_i(v_i)$. By the same independence, the second factor at $k=g_i(v_i)$ gives a version of the law of $v_{-i}$ given $(s,v_i)$. For the conditional law used in \eqref{eq:incentive} and \eqref{eq:participation}, we have
\[
\Law(v_{-i}\mid s,v_i)=\Law\bigl(v_{-i}\mid s_{-i}=(s-g_i(v_i))\bmod1\bigr)
\]
almost surely.

To apply IC and IR at every atom, let $G$ be the measurable subset of $A_i$ on which this measure identity holds. Thus IC holds for every report and IR holds on $G$, and $G$ has probability one. Since $(s,v_i)\in G$ almost surely and $g_i(v_i)=k_i$, conditioning on $(s,k_i)$ with marginal law $Q_i(k)$ gives $Q_i(k)(\{z:(s,z)\in G,\ g_i(z)=k\})=1$ for almost every $(s,k)$.

Fix a pair $(s,k)$ for which $Q_i(k)$ is countably supported, the product law and the preceding probability-one statement hold, and the conditional expected absolute transfer is finite. Write $Q=Q_i(k)$ and $H=\Law(v_{-i}\mid s_{-i}=(s-k)\bmod1)$, and let $T=\{z\ge0:Q(\{z\})>0\}$. Since $Q$ is countably supported, its atoms in $T$ carry all its mass. The set in the preceding probability-one statement has $Q$-mass one, and each atom has positive mass, so every $z\in T$ satisfies $(s,z)\in G$ and $g_i(z)=k$. Thus every type in $T$ faces the same law $H$ of the other buyers' truthful reports.

By the product law,
\[
\sum_{z\in T} Q(\{z\})\int |t_i(s,(z,w))|\,H(dw)
=\Exp[|t_i(s,v)|\mid s,k_i=k]<\infty.
\]
Thus the transfer at each report $z\in T$ is $H$-integrable because $Q(\{z\})>0$. Let $\bar x(z)$ and $\bar t(z)$ be the allocation and transfer at report $z$ averaged over $H$; then $\sum_{z\in T}Q(\{z\})|\bar t(z)|<\infty$. Since IC holds for every report at each true type $z\in T$, it applies in particular to any deviation report $z'\in T$. Using the common law $H$, IC and IR become, for all $z,z'\in T$,
\[
z\bar x(z)-\bar t(z)\ge z\bar x(z')-\bar t(z'),
\qquad z\bar x(z)-\bar t(z)\ge0.
\]
We have reduced the conditional problem to ordinary one-buyer IC and IR.

\textit{The revenue upper bound.}
A countable set need not admit an increasing enumeration, so start with a finite subset. Order any nonempty finite subset of $T$ as $z_1<\cdots<z_m$. Adding the IC inequalities in both directions between $z_{j-1}$ and $z_j$ gives
\[
(z_j-z_{j-1})[\bar x(z_j)-\bar x(z_{j-1})]\ge0.
\]
Set $\Delta_1=\bar x(z_1)$ and $\Delta_j=\bar x(z_j)-\bar x(z_{j-1})$ for $j\ge2$. These increments are nonnegative. IR at $z_1$ gives $\bar t(z_1)\le z_1\Delta_1$, and the IC constraint for type $z_j$ against reporting $z_{j-1}$ gives $\bar t(z_j)-\bar t(z_{j-1})\le z_j\Delta_j$. Summing yields $\bar t(z_j)\le\sum_{h=1}^j z_h\Delta_h$. It follows that
\[
\begin{aligned}
\sum_{j=1}^m Q(\{z_j\})\bar t(z_j)
&\le \sum_{j=1}^m Q(\{z_j\})\sum_{h=1}^j z_h\Delta_h\\
&=\sum_{h=1}^m z_h\Delta_h Q(\{z_h,\ldots,z_m\})\\
&\le \sum_{h=1}^m z_h\Delta_h Q([z_h,\infty))\\
&\le \rho(Q)\sum_{h=1}^m\Delta_h
=\rho(Q)\bar x(z_m)\le \rho(Q).
\end{aligned}
\]
If $T$ is finite, take the finite subset to be all of $T$.

If $T$ is infinite, take finite sets $T_m\uparrow T$. Applying the finite-subset bound to each $T_m$ with its elements ordered by value gives $\int\bar t(z)\mathbf{1}_{T_m}(z)\,Q(dz)\le\rho(Q)$ for every $m$. The truncated transfers converge to $\bar t$ $Q$-almost surely and are bounded in absolute value by the absolute value of the averaged transfer, which is integrable under $Q$. By dominated convergence, $\int\bar t(z)\,Q(dz)=\lim_{m\to\infty}\int\bar t(z)\mathbf{1}_{T_m}(z)\,Q(dz)\le\rho(Q)$.

Thus
\[
\Exp[t_i(s,v)\mid s,k_i=k]=\sum_{z\in T}Q(\{z\})\bar t(z)\le\rho(Q).
\]
Since $s\perp v_i$ and $k_i=g_i(v_i)$, each key is uniform conditional on $s$, so averaging the preceding almost-sure bound over the key and summing over buyers gives
\begin{equation}\label{eq:prop1-revenue-bound}
\Exp\!\left[\sum_{i=1}^n t_i(s,v)\,\middle|\,s\right]
\le\sum_{i=1}^n\int_0^1\rho(Q_i(k))\,dk
\end{equation}
for almost every $s$.

\textit{Attainment by posted prices.}
Under truthful reporting $\kappa_i=k_i$, and conditional on $(s,k_i=k)$ the purchase probability is $Q_i(k)([p_i(k),\infty))$. Since $p_i(k)$ is a monopoly price, buyer $i$'s conditional expected payment is $\rho(Q_i(k))$ for almost every $(s,k)$. Integrating over each uniform key conditional on $s$ shows that the mechanism attains the upper bound in \eqref{eq:prop1-revenue-bound} and is seller-optimal.
\end{proof}

\subsection{Keys for the three vertices}

Work first with one buyer under the distributional assumptions in Section~\ref{sec:model}. Write $F$ for the continuous CDF of her value $v$, $p^*$ for her unique monopoly price, and $r=p^*[1-F(p^*)]>0$ for the seller's monopoly revenue from that buyer. For a nondecreasing function $C$ with domain $D$, define $C^{-1}(y)=\inf\{v\in D:C(v)\ge y\}$.

The three lemmas construct keys for the full-extraction vertex $\welfarepoint{1}$ and the two minimum-revenue vertices: $\welfarepoint{2}$ with zero consumer surplus and $\welfarepoint{3}$ with efficient trade.

\begin{samepage}
\begin{lemma}\label{lem:full}
There are a uniform key $k=g(v)$ and a measurable price function $p\colon[0,1)\to[0,\infty)$ such that, for almost every $k\in[0,1)$, the conditional distribution $\Law(v\mid k)$ is a point mass at the monopoly price $p(k)=F^{-1}(k)$. Posting $p(k)$ yields expected revenue $\Exp[v]$ and zero consumer surplus.
\end{lemma}
\end{samepage}

\begin{proof}
Let $U$ be uniform on $(0,1)$. Right-continuity of the CDF gives $F^{-1}(U)\le x$ if and only if $U\le F(x)$, so $F^{-1}(U)$ has law $F$. Continuity of $F$ gives $F(F^{-1}(y))=y$ for every $y\in(0,1)$. Thus $F(v)$ is uniform and $F^{-1}(F(v))=v$ almost surely, because $v$ has the same law as $F^{-1}(U)$. Take $g(v)=F(v)$ except that $g(v)=0$ when $F(v)=1$. The replacement on $\{F(v)=1\}$ affects only a null event, so the key is uniform on $[0,1)$. The price function $p(k)=F^{-1}(k)$ is measurable because it is nondecreasing. Since $v=p(k)$ almost surely, the conditional distribution given $k$ is a point mass at $p(k)$. Posting $p(k)$ maximizes revenue: the buyer purchases at equality, a lower price earns less, and a higher price earns zero. Thus revenue is $v$ and consumer surplus is zero almost surely.
\end{proof}

\begin{samepage}
\begin{lemma}\label{lem:zeroexact}
There are a uniform key $k=g(v)$ and a measurable price function $p\colon[0,1)\to[0,\infty)$ such that, for almost every $k\in[0,1)$, the conditional distribution $\Law(v\mid k)$ is countably supported and $p(k)$ is a monopoly price equal to the largest valuation in its set of atoms. Posting $p(k)$ yields expected revenue $r$ and zero consumer surplus.
\end{lemma}
\end{samepage}

\begin{samepage}
\begin{lemma}\label{lem:efficientexact}
There are a uniform key $k=g(v)$ and a measurable price function $p\colon[0,1)\to[0,\infty)$ such that, for almost every $k\in[0,1)$, the conditional distribution $\Law(v\mid k)$ is countably supported and $p(k)$ is a monopoly price no greater than any valuation in its set of atoms. Posting $p(k)$ yields expected revenue $r$ and expected consumer surplus $\Exp[v]-r$.
\end{lemma}
\end{samepage}

Appendix~\ref{sec:exact} proves Lemmas~\ref{lem:zeroexact} and~\ref{lem:efficientexact}.

\subsection{Proof of Theorem~\ref{thm:exact}}

\begin{proof}
\textit{Attaining the vertices.} For each vertex, let $g_i$ and $p_i$ be buyer $i$'s key and price functions obtained by applying the corresponding one-buyer lemma to $F_i$. For $\welfarepoint{1}$, use Lemma~\ref{lem:full}. With the modular signal, Proposition~\ref{prop:keyconversion} shows that posting the recovered values as prices is seller-optimal. Each buyer trades and pays her value almost surely, so consumer surplus is zero and revenue is $w^*$.

\begin{samepage}
Apply Lemma~\ref{lem:zeroexact} to every buyer for $\welfarepoint{2}$ and Lemma~\ref{lem:efficientexact} to every buyer for $\welfarepoint{3}$. In each case, combine the keys through \eqref{eq:mod}. Both lemmas supply the measurable prices required by Proposition~\ref{prop:keyconversion}, so the posted-price mechanisms are seller-optimal. Adding the surplus pairs across buyers gives the desired vertex in each case.
\end{samepage}

\textit{Mixing the constructions.} Let $s_1,s_2,s_3$ be the modular signals for the constructions of $\welfarepoint{1}$, $\welfarepoint{2}$, and $\welfarepoint{3}$, respectively. The platform draws $J\in\{1,2,3\}$ independently of $v$ and publishes $(J,s_J)$ so the seller knows which pricing problem to solve. For any buyer $i$, index $j$ and Borel set $B$ of signal values,
\[
\Pr(J=j,\ s_J\in B\mid v_i)=\Pr(J=j)\Pr(s_j\in B),
\]
almost surely, because the $j$th construction preserves privacy. Thus the combined signal also preserves privacy.

For each $j$ with $\Pr(J=j)>0$, the joint distribution of $(v,s_j)$ conditional on $J=j$ is the same as when construction $j$ is used on its own, because $J$ is drawn independently of $v$. A competing feasible mechanism restricted to that component therefore faces the same conditional laws and retains the probability-one IC and IR constraints and transfer integrability. The competitor's conditional expected revenue on component $j$ is therefore no greater than that of the corresponding seller-optimal mechanism for almost every $s_j$. After observing $(J,s_J)$, the seller posts to each buyer $i$ the price prescribed by construction $J$ at $(s_J,\hat v_{-i})$. Each price is measurable and depends only on $(J,s_J,\hat v_{-i})$, so the combined rule is a feasible, seller-optimal posted-price mechanism. Expected surplus is the probability-weighted sum of the component outcomes, so by varying the probabilities of $J$ in its disclosure rule, the platform makes every convex combination of the three vertices attainable.

\textit{Bounds without privacy.} For any information structure, even without privacy restrictions, the seller can offer the fixed no-disclosure monopoly prices and ignore the signal. This mechanism is feasible and earns $\pi^*$ in expectation by iterated expectations. Seller optimality gives $\pi\ge\pi^*$, IR gives $u\ge0$, and the allocation bound $x_i(s,v)\le1$ gives $u+\pi=\Exp[\sum_i v_i x_i(s,v)]\le w^*$. All three bounds hold without the privacy constraint, and the constructions above attain the whole triangle under it, so the same surplus region is attainable with or without the constraint.
\end{proof}

\section{Discussion}\label{sec:discussion}

Our results show that a restriction prohibiting the disclosure of information about individual buyers' valuations does not necessarily limit the seller's ability to price discriminate. They also show that privacy constraints do not necessarily protect consumers. The key is the complementarity between the public signal and other buyers' reports. For any given buyer, the privacy constraint makes the public signal uninformative about that buyer's valuation. Independence of valuations makes the other buyers' truthful reports uninformative on their own. Together, however, the signal and these reports can reveal the buyer's valuation exactly. Our paper shows that such information complementarity makes price discrimination possible without directly disclosing individual buyers' valuations.

This information complementarity relies on the seller's ability to use other buyers' reports when setting prices. Suppose that each buyer's allocation and transfer could depend only on the public signal and her own report. Since the privacy constraint leaves each buyer's valuation distribution unchanged conditional on the public signal, the no-disclosure monopoly price would remain optimal. The uniqueness of that price implies that every seller-optimal mechanism of this form has the same allocation and payment as posting that price, almost surely. Thus the attainable welfare region would collapse to the single point $\welfarepoint{0}$. Thus, the entire surplus triangle remains attainable under the privacy constraint because the seller is allowed to use information obtained through other buyers' reports.

This distinction can also be understood in terms of the strength of the privacy constraint. Our paper requires the public signal to be independent of each buyer's valuation, but does not require the public signal to remain independent of a buyer's valuation after conditioning on all other buyers' valuations. Suppose instead that we imposed the stronger privacy constraint

\[
s\perp v_i\mid v_{-i}
\]

for every buyer. Under the product prior, this condition makes the probability of any signal event given the valuation vector unchanged by averaging over any one coordinate. Averaging over all coordinates in turn gives the unconditional probability of that event, so the signal is independent of the entire valuation vector. Consequently, the public signal could provide no additional information to the seller, and the attainable welfare region would collapse to the single point $\welfarepoint{0}$.

Finally, the independence of valuations is also important. When valuations are correlated, other buyers' reports may provide information about a buyer's valuation even without disclosure. For example, if all buyers share the same realized valuation almost surely, the seller can identify each buyer's valuation exactly from the other buyers' truthful reports. Full surplus extraction is then possible under any disclosure rule, and the attainable welfare region collapses to a single point. More generally, when valuations are correlated, the seller may be able to earn revenue exceeding the sum of the monopoly revenues based on the marginal distributions. Therefore, the triangle defined by the marginal distributions need not characterize the attainable welfare region.

\clearpage
\appendix
\numberwithin{lemma}{section}
\renewcommand{\thelemma}{\thesection\arabic{lemma}}
\section{Proofs of the Key Constructions}\label{sec:exact}

In the constructions below, set $g(v)=0$ wherever its formula is undefined or falls outside $[0,1)$, a Borel set of valuations of $F$-measure zero. Likewise, set $p(k)=0$ wherever its formula is undefined or falls outside $[0,\infty)$, a Borel set of keys of Lebesgue measure zero.

\subsection{A key with constant conditional monopoly revenue}

\begin{lemma}\label{lem:neutral}
If $v\ge0$ has a continuous CDF $F$, finite mean, monopoly revenue $r>0$, and unique monopoly price $p^*$, then there exists a measurable key $k=g(v)$, uniform on $[0,1)$, such that, for almost every key $k$, the conditional distribution $\Law(v\mid k)$ is countably supported and has monopoly revenue $r$, attained at $p^*$. In constructing the key, the partition above $p^*$ can start at any finite $b_0>p^*$.
\end{lemma}

\begin{proof}
\textit{Partition the value space.} Preserving each interval’s mass will bound revenue at prices in $(\underline a,\bar a)$ by $\bar a[1-F(\underline a)]$. To keep this bound at $r$, construct intervals by setting the upper endpoint for a given lower endpoint $\underline a$ to
\[
\bar a\coloneqq\frac{r}{1-F(\underline a)}.
\]
Each resulting interval satisfies
\begin{equation}\label{eq:foldinterval}
\bar a[1-F(\underline a)]\le r.
\end{equation}

Below $p^*$, start at $a_0\coloneqq0$ and set $a_{j+1}=r/[1-F(a_j)]$ while $a_j<p^*$. Then $a_j<a_{j+1}\le p^*$ because $a_j[1-F(a_j)]<r=p^*[1-F(p^*)]$ and $F(a_j)\le F(p^*)$. Above $p^*$, choose any finite starting point $b_0>p^*$. The freedom to choose $b_0$ lets us make the cutoff in the next proof a partition endpoint. For the upward steps, set $c_0=b_0$ and $c_{j+1}=r/[1-F(c_j)]$ while $F(c_j)<1$. Uniqueness of $p^*$ gives $c_{j+1}>c_j$. For the downward steps from $b_0$, set
\[
b_{j+1}\coloneqq F^{-1}\!\left(1-\frac r{b_j}\right).
\]
Uniqueness of $p^*$ gives $F(p^*)=1-r/p^*<1-r/b_j<F(b_j)$ whenever $b_j>p^*$. Continuity then gives $p^*<b_{j+1}<b_j$ and $F(b_{j+1})=1-r/b_j$. Any finite limit $c$ of an infinite endpoint sequence satisfies $c[1-F(c)]=r$ by the endpoint recursion and continuity. Thus $c=p^*$. The sequences of $a_j$ and $b_j$ each reach or converge to $p^*$. The sequence of $c_j$ either stops at a point with no upper-tail mass or diverges to infinity.

Both endpoint rules give equality in \eqref{eq:foldinterval}. The countably many endpoints and their possible accumulation point $p^*$ have zero mass, as does a tail beyond a stopping endpoint. Thus the intervals between consecutive endpoints cover $F$-almost every value. Index the positive-probability intervals by $I_j=(\alpha_j,\beta_j)$.

\textit{Use the rank within each interval.} For $v\in I_j$, define
\begin{equation}\label{eq:foldkey}
g(v)\coloneqq\frac{F(v)-F(\alpha_j)}{F(\beta_j)-F(\alpha_j)}.
\end{equation}
The rank in \eqref{eq:foldkey} is uniform conditional on $v\in I_j$. Thus for every Borel set $S\subset(0,1)$,
\[
\Pr(v\in I_j,\ g(v)\in S)=\bigl[F(\beta_j)-F(\alpha_j)\bigr]\int_S dk.
\]
Summing over intervals shows that the key is uniform and independent of the index of the interval containing $v$. On $v\in I_j$, the generalized inverse recovers the value almost surely as
\[
F^{-1}\!\big((1-k)F(\alpha_j)+kF(\beta_j)\big).
\]
For $0<k<1$ the cumulative level lies strictly between $F(\alpha_j)$ and $F(\beta_j)$, so the recovered value lies in $I_j$. To compute the joint probability with the additional restriction that $v$ lie in a Borel set $T$, restrict $S$ to keys for which the recovered value lies in $T$. The resulting joint probabilities identify the conditional law at almost every key: the atom in $I_j$ has probability $F(\beta_j)-F(\alpha_j)$.

\textit{Check optimality.} Fix a key $k$ for which this description of the conditional distribution holds. For any price $p$ in an interval $(\underline a,\bar a)$ of the original partition, including one with zero probability, no conditional atom lies at an endpoint, and $\Pr(v\ge\underline a\mid k)=1-F(\underline a)$. Hence
\[
p\Pr(v\ge p\mid k)\le\bar a[1-F(\underline a)]\le r.
\]

For an endpoint $p>0$, take $p_m\uparrow p$ through interval interiors and use $\Pr(v\ge p_m\mid k)\downarrow\Pr(v\ge p\mid k)$. Revenue is zero at $p=0$ and beyond any stopping endpoint with zero tail.

Since no interval crosses $p^*$, $p^*\Pr(v\ge p^*\mid k)=p^*[1-F(p^*)]=r$.
\end{proof}

\subsection{A key attaining zero consumer surplus at minimum revenue}

\begin{proof}[Proof of Lemma~\ref{lem:zeroexact}]
\textit{Choose the cutoffs and mixture weights.} To obtain zero consumer surplus, we construct the key so that the seller charges the buyer her full value whenever she receives the good. To obtain revenue $r$, the construction serves exactly an upper tail whose total value is $r$. Choose its cutoff $b$ by
\begin{equation}\label{eq:zero_tail}
\int_{[b,\infty)}v\,dF(v)=r.
\end{equation}
Take the smallest such cutoff. A finite $b>p^*$ exists: the tail-value integral is continuous, tends to zero, and at $p^*$ exceeds $r$ by the positive integral $\int_{[p^*,\infty)}(v-p^*)\,dF(v)$. That tail has mass $r/p^*>0$ and no atom at $p^*$. Set $L=1-r/b$. Since $p^*$ is the unique monopoly price and $b>p^*$,
\[
F(p^*)=1-\frac r{p^*}<L<F(b).
\]

We choose $v_0$ so that the prior has mass $r/b$ above it. With that same tail mass in each conditional distribution, prices strictly between $v_0$ and $b$ earn at most $r$. Start the partition in Lemma~\ref{lem:neutral} at $b_0=b$. The first downward step gives the endpoint $v_0=F^{-1}(L)$, which lies in $(p^*,b)$ and satisfies $F(v_0)=L$ by continuity.

Assign probability $r/x$ to a top value $x\ge b$ in each conditional distribution, so that selling only to a buyer with value $x$ earns $r$. Keep the original masses of the lower intervals $I_j$, defined below, totaling $L$. Assign the remaining probability $r/b-r/x$ to a single value in $(v_0,b)$. The three probabilities sum to one.

\textit{Define the key.} For $x\ge b$, define the cumulative functions
\[
B(x)\coloneqq\frac1r\int_{[b,x]}v\,dF(v),\qquad
M(x)\coloneqq\frac rb B(x)-F(x)+F(b).
\]
Both functions are continuous and nondecreasing, with increments
\[
dB(x)=\frac xr dF(x),\qquad dM(x)=\left(\frac xb-1\right)dF(x).
\]
Moreover, $B(\infty)=1$, while $M(\infty)=F(b)-L$ is the probability of the middle region $(v_0,b)$.

Use the positive-probability intervals $I_j=(\alpha_j,\beta_j)$ of this partition below $v_0$. Because $v_0$ is already an endpoint, no interval needs to be cut. Define the key directly by
\begin{equation}\label{eq:zero_recipe}
g(v)\coloneqq
\begin{cases}
\dfrac{F(v)-F(\alpha_j)}{F(\beta_j)-F(\alpha_j)},&v<v_0,\ v\in I_j,\\[6pt]
B\!\left(M^{-1}(F(v)-L)\right),&v_0<v<b,\\[3pt]
B(v),&v\ge b,\ B(v)<1.
\end{cases}
\end{equation}

\textit{Verify uniformity and the conditional laws.} Put $p(k)=B^{-1}(k)$ and take Borel sets $S\subset(0,1)$ and $T\subset[0,\infty)$. The continuous cumulative function $B$ sends the atomless measure $dB=(v/r)dF(v)$ to Lebesgue measure $dk$ on $(0,1)$, and $B^{-1}(B(v))=v$ for $dB$-almost every $v$. Undoing the weight $v/r$ gives
\[
\Pr(v\ge b,\ g(v)\in S,\ v\in T)=\int_S \one\{p(k)\in T\}\frac{r}{p(k)}\,dk.
\]
On $v_0<v<b$, the unnormalized law of $F(v)-L$ is Lebesgue measure on $(0,F(b)-L)$. Therefore $x=M^{-1}(F(v)-L)$ has unnormalized law $dM$, with $M(x)=F(v)-L$ almost surely. Since $dM=(r/b-r/x)dB$ is absolutely continuous with respect to $dB$, $B^{-1}(B(x))=x$ almost surely under $dM$ as well. With $k=B(x)$ we thus have $x=p(k)$ and the original value is $F^{-1}(L+M(p(k)))$ almost surely. Applying $B$ to the law of $x$ now gives
\[
\Pr(v_0<v<b,\ g(v)\in S,\ v\in T)
=\int_S \one\{F^{-1}(L+M(p(k)))\in T\}\left(\frac rb-\frac{r}{p(k)}\right)\,dk.
\]
Continuity gives $B(p(k))=k$ for $0<k<1$, implying positive $F$-mass in both $(b,p(k)]$ and $(p(k),\infty)$, so $p(k)>b$. Since $dM(x)=(x/b-1)dF(x)$, we have $0<M(p(k))<F(b)-L$. Thus the middle atom lies strictly in $(v_0,b)$ and has positive weight $r/b-r/p(k)$. Each lower interval contributes its original probability times the Lebesgue measure of the keys in $S$ for which the value recovered in Lemma~\ref{lem:neutral} lies in $T$. Taking $T=[0,\infty)$, the densities sum to one, so $k$ is uniform. There is one atom per lower interval and two above $v_0$, so the conditional distributions are countably supported. The price $p(k)$ is measurable.

\textit{Verify price optimality and payoffs.} For almost every key $k$, the largest valuation in the conditional distribution is $p(k)$ with probability $r/p(k)$, so posting $p(k)$ earns $r$ and gives zero consumer surplus.

For $q<v_0$, each lower interval $I_j$ retains its original mass and the total mass above $v_0$ is $1-F(v_0)=r/b$. Thus the conditional tail mass above the left endpoint $\alpha_j$ of any lower interval remains $1-F(\alpha_j)$. The bound in Lemma~\ref{lem:neutral} then applies, including at the endpoints by taking prices from below as in that lemma. For the remaining prices,
\[
\begin{aligned}
q\Pr(v\ge q\mid k)&\le q\,r/b\le r &&(v_0\le q<b),\\
q\Pr(v\ge q\mid k)&=q\frac{r}{p(k)}\one\{q\le p(k)\}\le r &&(q\ge b).
\end{aligned}
\]

Since no lower interval $I_j$ crosses $p^*$ and the original mass above $v_0$ is preserved, $p^*\Pr(v\ge p^*\mid k)=r$ as well.
\end{proof}

\subsection{A key attaining efficiency at minimum revenue}

\begin{proof}[Proof of Lemma~\ref{lem:efficientexact}]
\textit{Choose the cutoff and mixture weights.} For a candidate cutoff $a\in[0,p^*]$ and a low valuation $0\le\ell<a$, let $m\coloneqq1-F(a)$; assigning probability $m\ell/r$ to the high-value component constructed below makes its contribution to revenue at $p^*$ equal to $\ell$, the revenue from selling with probability one at price $\ell$.

To preserve the original low-value mass $dF(\ell)$, the measure of keys indexed by their associated low values must have mass element $dF(\ell)/(1-m\ell/r)$: each key paired with $\ell$ assigns only the fraction $1-m\ell/r$ to that value. Thus for $0\le a\le p^*$ the total probability needed for keys with a low value is
\begin{equation}\label{eq:Afunction}
A(a)\coloneqq\int_{[0,a)}\frac{dF(v)}{1-v[1-F(a)]/r}.
\end{equation}
For $a<p^*$, the denominator is strictly positive on $[0,a)$ because $a[1-F(a)]<r$. For $a_0\in(0,p^*)$, take $0<\epsilon<\min\{a_0,p^*-a_0\}$ small enough that $(a_0+\epsilon)[1-F(a_0-\epsilon)]<r$. This gives a common positive lower bound for the denominators on $v<a$ when $|a-a_0|<\epsilon$. The integrands, extended by zero for $v\ge a$, converge pointwise except at $v=a_0$, which has zero probability, so dominated convergence gives continuity. As $a$ approaches zero, $A(a)\le F(a)/(1-a/r)$ for $a<r$, which tends to zero.

Define $A(p^*)$ by the possibly infinite integral over $v<p^*$. As $a\uparrow p^*$ the extended integrands converge at every $v<p^*$, so Fatou's lemma gives $\liminf_{a\uparrow p^*}A(a)\ge A(p^*)$. Choose $a=p^*$ if $A(p^*)\le1$: at almost every unpaired key, posting $p^*$ then results in trade with probability one. Otherwise choose a lower cutoff so that keys paired with low values use all available mass. Fatou's bound gives some $c<p^*$ with $A(c)>1$. By continuity the solutions to $A(a)=1$ in $[0,c]$ form a nonempty closed set, so choose its smallest element. The smallest solution is positive because $A(a)\to0$ as $a\downarrow0$.

\textit{Construct the conditional mixtures.} For the chosen cutoff, put $m\coloneqq1-F(a)$ and $\gamma\coloneqq m/r$. Then $\gamma a\le1$. Let $H\coloneqq\Law(v\mid v\ge a)$. Its monopoly revenue is $r/m=1/\gamma$, and its unique monopoly price is still $p^*$. Prices at least $a$ yield their original revenues divided by $m$, and a price $p<a$ earns $p<a\le r/m$. The distribution $H$ is atomless, so Lemma~\ref{lem:neutral} gives a uniform key $h(v)$ under $H$. For almost every rank $\eta$, the conditional law given $h(v)=\eta$ under $H$ has monopoly revenue $1/\gamma$, attained at $p^*$.

For each low value $\ell<a$, consider a conditional distribution of the form
\begin{equation}\label{eq:two_layer}
(1-\gamma\ell)\delta_\ell+\gamma\ell \widetilde H_\ell.
\end{equation}
Here $\delta_\ell$ is the point mass at $\ell$, and $\widetilde H_\ell$ is one of the conditional laws $\Law(v\mid v\ge a,\,h(v)=\eta)$ for a rank $\eta$ with these properties. The rank is assigned below. At any price $q\le\ell$, the buyer purchases with probability one, so revenue is at most $\ell$, with equality at $\ell$. For $q>\ell$, only the high-value component can generate sales, and its contribution to revenue satisfies
\[
\gamma\ell\,q\,\widetilde H_\ell([q,\infty))\le\gamma\ell\,\rho(\widetilde H_\ell)=\ell.
\]

Equality also holds at $p^*$ because $p^*\ge a>\ell$ and $p^*$ maximizes revenue under $\widetilde H_\ell$.

\textit{Define the key.} For $0\le v\le a$, put
\[
E(v)\coloneqq\int_{[0,v)}\frac{dF(w)}{1-\gamma w},\qquad
D(v)\coloneqq E(v)-F(v)=\int_{[0,v)}\frac{\gamma w}{1-\gamma w}dF(w).
\]
Thus $E(a)=A(a)$ and $D(a)=A(a)-F(a)\le m$. The function $D(v)$ gives the high-value mass needed to accompany low values below $v$. On the matched high-value branch, $D^{-1}$ maps the unnormalized high-value rank to the corresponding low value, and $E$ assigns a key to the corresponding low value. Using $h$, define
\begin{equation}\label{eq:efficient_recipe}
g(v)\coloneqq
\begin{cases}
E(v),&v<a,\ E(v)<1,\\[3pt]
E\!\left(D^{-1}(m h(v))\right),&v\ge a,\ m h(v)<D(a),\\[3pt]
m h(v)+F(a),&v\ge a,\ m h(v)\ge D(a).
\end{cases}
\end{equation}

\textit{Verify uniformity and the conditional laws.} Set $\eta=h(v)$ on the high region and $\ell(k)=E^{-1}(k)$ on $0<k<A(a)$. Write $H_\eta=\Law(v\mid v\ge a,\ h(v)=\eta)$ for the conditional law from Lemma~\ref{lem:neutral}. Since $E(a)=A(a)\le1$ even when $a=p^*$, $E$ and $D$ are continuous cumulative functions of finite atomless measures. Thus $0<\ell(k)<a$ on $(0,A(a))$. The continuous cumulative function $E$ sends $dE$ to Lebesgue measure $dk$ on $(0,A(a))$, and $E^{-1}(E(v))=v$ for $dE$-almost every $v$. Hence for Borel $S\subset(0,A(a))$ and $T\subset[0,\infty)$, using $dF=(1-\gamma v)dE$ gives
\[
\Pr(v<a,\ g(v)\in S,\ v\in T)
=\int_S\one\{\ell(k)\in T\}(1-\gamma\ell(k))\,dk.
\]
If $D(a)>0$, the unnormalized law of $m\eta$ on the matched high-value branch is Lebesgue measure on $(0,D(a))$. The map $D^{-1}$ sends this measure to $dD=\gamma v\,dE$, which $E$ then sends to $\gamma\ell(k)\,dk$. Since $D(D^{-1}(m\eta))=m\eta$ and recovery by $E^{-1}$ holds also under $dD$ by absolute continuity with respect to $dE$, the original rank is $\eta=D(\ell(k))/m$ almost surely on the matched branch. Conditioning on the recovered high-value rank gives
\[
\Pr(v\ge a,\ g(v)\in S,\ v\in T)
=\int_S\gamma\ell(k)\,H_{D(\ell(k))/m}(T)\,dk.
\]
The map $k\mapsto D(\ell(k))/m$ sends $\gamma\ell(k)\,dk$ to $m\,d\eta$ on $(0,D(a)/m)$. Since $\gamma\ell(k)>0$ on $(0,A(a))$, properties valid for almost every high rank therefore hold for almost every key in this interval. If $D(a)=0$ the branch is empty. Taking $T=[0,\infty)$ shows the two contributions have total density one on $(0,A(a))$; for general $T$ they give \eqref{eq:two_layer} with $\widetilde H_{\ell(k)}=H_{D(\ell(k))/m}$ for almost every matched key $k$.

On the residual branch, $k=m\eta+F(a)$ by \eqref{eq:efficient_recipe}. As $m\eta$ runs from $D(a)$ to $m$, $k$ runs from $A(a)$ to $1$, and the mass of the branch is $1-A(a)$. For Borel $S\subset(A(a),1)$, conditioning on $\eta$ and substituting $dk=m\,d\eta$ gives the joint probability of $g(v)\in S$ and $v\in T$ as $\int_S H_{(k-F(a))/m}(T)\,dk$. For $T=[0,\infty)$, the integral gives the length of $S$, completing the proof that $k$ is uniform.

\textit{Verify price optimality and payoffs.} If $A(a)<1$, then $a=p^*$. The residual conditional distributions are the conditional laws of $H$ given $h(v)$ obtained above from Lemma~\ref{lem:neutral}: for almost every residual key they are countably supported on $[a,\infty)$ and admit $a$ as a monopoly price. For almost every matched key, the conditional distribution is also countably supported: it adds one low atom to a countably supported conditional law of $H$. Define the measurable price $p(k)$ by $p(k)=E^{-1}(k)$ for $0<k<A(a)$ and $p(k)=a$ for $A(a)\le k<1$ on the residual branch. For almost every key, $p(k)$ is a monopoly price no greater than any valuation in the conditional distribution's set of atoms.

The original price $p^*$ remains optimal for almost every key. Hence
\begin{equation}\label{eq:tower}
\Exp[\rho(\Law(v\mid k))]
=\Exp[p^*\Pr(v\ge p^*\mid k)]
=p^*\Pr(v\ge p^*)=r.
\end{equation}
The buyer receives a unit almost surely at price $p(k)$, so expected allocated value is $\Exp[v]$ and expected consumer surplus is $\Exp[v]-r$.
\end{proof}

\begingroup
\setstretch{1.0}
\PaperBibliography
\endgroup

\end{document}